\documentclass[11pt]{diazessay} 
\usepackage[pagebackref=true, hidelinks, pdftex, pdfkeywords={paglen, crawford, excavating ai, jaffe, affective computing, machine learning}]{hyperref}
\usepackage[utf8]{inputenc}
\usepackage[T1]{fontenc}    
\usepackage{CJKutf8} 
\usepackage{multirow}
\hypersetup{
  colorlinks   = true, 
  urlcolor     = blue, 
  linkcolor    = blue, 
  citecolor   = cyan 
}
\newcommand{\q}[1]{``#1''}

\title{\textbf{\q{Excavating AI} Re-excavated: \\Debunking a Fallacious Account of the \\JAFFE Dataset}}

\author{\textbf{Michael J. Lyons} \\ \textit{Ritsumeikan University}} 

\date{}

\begin{document}
\begin{CJK}{UTF8}{min}

\maketitle 

\begin{abstract}
\noindent 
Twenty-five years ago, my colleagues Miyuki Kamachi and Jiro Gyoba and I designed and photographed JAFFE, a set of facial expression images intended for use in a study of face perception. In 2019, without seeking permission or informing us, Kate Crawford and Trevor Paglen exhibited JAFFE in two widely publicized art shows. In addition, they published a nonfactual account of the images in the essay \q{Excavating AI: The Politics of Images in Machine Learning Training Sets.} The present article recounts the creation of the JAFFE dataset and unravels each of Crawford and Paglen's fallacious statements. I also discuss JAFFE more broadly in connection with research on facial expression, affective computing, and human-computer interaction.
\end{abstract}
\hspace*{3.6mm}\textit{Keywords:} facial expression, machine learning, training sets,  \\ \hspace*{20mm} affective computing,  artificial intelligence  


\vspace{30pt}
\begin{quote}
\q{To make a mistake and not correct it: this is a \emph{real} mistake.}
\begin{flushright}\emph{Confucius, The Analects}\end{flushright}
\end{quote}
\noindent
In September of 2019, Kate Crawford and Trevor Paglen opened the art exhibition \href{https://www.fondazioneprada.org/project/training-humans/}{\emph{Training Humans}} at the Milan Osservatorio of the Fondazione Prada. The stated objective of the exhibition was to explore:
\begin{quote}
\textit{
$\ldots$ how humans are represented, interpreted and codified through training datasets, and how technological systems harvest, label and use this material.
}
\end{quote}
Accompanying the exhibition, the essay \q{Excavating AI: The Politics of Images in Machine Learning Training Sets} was self-published online.\footnote{\url{https://excavating.ai/}} To our surprise and dismay, \emph{Training Humans} exhibited more than a hundred facial images taken from the JAFFE dataset,\footnote{JAFFE is available for non-commercial scientific research via the \href{https://zenodo.org/record/3451524}{Zenodo repository.} Note that all JAFFE images in this article are subject to specific terms of use and may not be reused without permission, regardless of the license applied to the document as a whole.} a set of posed facial expressions photographed by my colleagues and me for use in our research \cite{lyons1998codingivc}. Though we provide the JAFFE dataset to other researchers for purposes of non-commercial scientific research, Crawford and Paglen's public exhibition of the images was a violation of the terms of use \cite{lyons2020excavating}. 
Meanwhile, \q{Excavating AI} spotlighted JAFFE as an `anatomical model' of machine learning training data, using it to expound an `archaeological method' of \q{cataloguing the principles and values by which [it] was constructed.}
The essay so misportrayed JAFFE as to render it nearly unrecognizable by those most familiar with the dataset: its creators.

A previous commentary \cite{lyons2020excavating} outlined the errors in Crawford and Paglen's account of JAFFE and explained that their unauthorized use of the images violated not only the terms of use but also the principle of informed consent. By misusing JAFFE, the authors of \q{Excavating AI} breached the very ethics they claimed to espouse.  Despite the objections raised in my commentary,  they continue to propagate a fallacious account of JAFFE by republishing the nearly unmodified text of \q{Excavating AI}---it has now been reprinted in a publication\footnote{\url{https://www.biennial.com/journal/issue-9/}} accompanying the 11th edition of the Liverpool Biennial of Contemporary Art \cite{krysa2021stages} and published in a special issue of the peer-reviewed scholarly journal \emph{AI \& Society} \cite{azar2021springer}.\footnote{An editor of the special issue assured me that the latter publication had indeed undergone peer review, but the text of the \emph{AI \& Society} version is nearly identical to the original web essay, casting doubt on the thoroughness of the referees' evaluations. As of June/July 2021, the article has been retracted from the \emph{AI \& Society} website but is still available elsewhere online.} This has motivated the present in-depth exposition of Crawford and Paglen's misleading statements about JAFFE. 

The present article describes the JAFFE dataset in some detail, summarizing information from the published documentation of the dataset. Each of Crawford and Paglen's claims regarding JAFFE is explicitly refuted. Analysis of the general pattern of the claims shows that \q{Excavating AI} makes liberal use of the common fallacy of informal logic known as the \emph{Straw Man} argument \cite{walton2008informal}. Concurrently, the article situates JAFFE in the broader context of experimental psychology, neural computation, affective computing, and human-computer interaction.

\section*{What is JAFFE?}
\label{jaffewhat}
JAFFE---The Japanese Female Facial Expression dataset---is a set of images depicting facial expressions posed by Japanese women, accompanied by semantic ratings on nouns describing the expressions. JAFFE was created specifically for use in an experiment that compared a biologically-inspired representation of facial appearance \cite{lyons1996model,lyons1997gabor,lyons2000linked} with facial expression perception by human observers \cite{lyons1998coding,lyons1997gabor2,lyons1997v1, lyons1998codingivc}. JAFFE is described as a dataset because it includes observer ratings for each facial expression image. Without this data, it would be more accurate to describe JAFFE as an image set or stimulus set.

\begin{figure}[t]
\begin{center}
	\includegraphics[width=1.\linewidth]{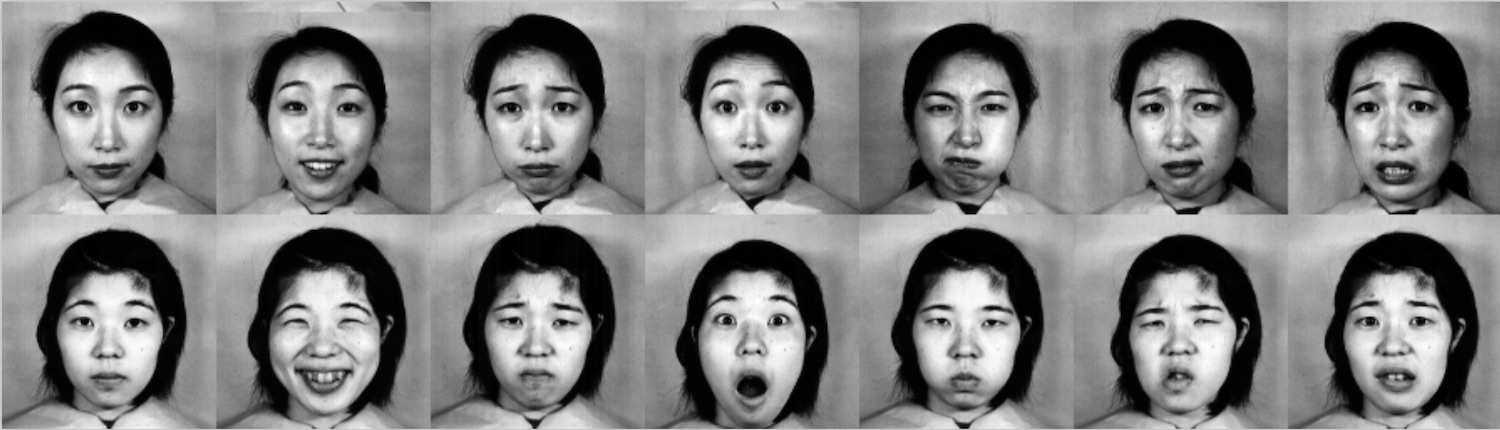}
	\caption{Posed facial expression images from the JAFFE dataset.}
\label{jaffe}
\end{center}
\end{figure}

We planned and assembled JAFFE in 1996 in collaboration with Miyuki Kamachi and Jiro Gyoba of the Psychology Department at Kyushu University, Japan. Miyuki recruited volunteers in the Psychology Department and asked them to pose three or four examples of six facial expressions (happiness, sadness, surprise, anger, disgust, fear) and a neutral or expressionless face.\footnote{Facial expression poses were requested using the Japanese adjectival nouns: 幸福、\newline 悲しみ、驚き、怒り、嫌悪、恐れ、and　無表情　for the expressionless pose. We used the same adjectival nouns to obtain semantic ratings by Japanese female observers.} These six expressions are the \emph{basic facial expressions} (BFE) studied by Paul Ekman and many others and thought by some of these scientists to be recognizable, to an extent, in all cultures \cite{ekman1992argument}.\footnote{The universality hypothesis for facial expression is attributed to Charles Darwin, who thought that at least seven categories of facial expressions---the six BFE and additionally contempt---are recognizable by all humans \cite{darwin1872express}. The extent to which facial expressions are universal remains a matter of controversy \cite{barrett2021ai,barrett2019emotional} and a topic of active research \cite{cowen2021sixteen,srinivasan2018cross}.}

\begin{figure}[t]
\begin{center}
	\includegraphics[width=1.\linewidth]{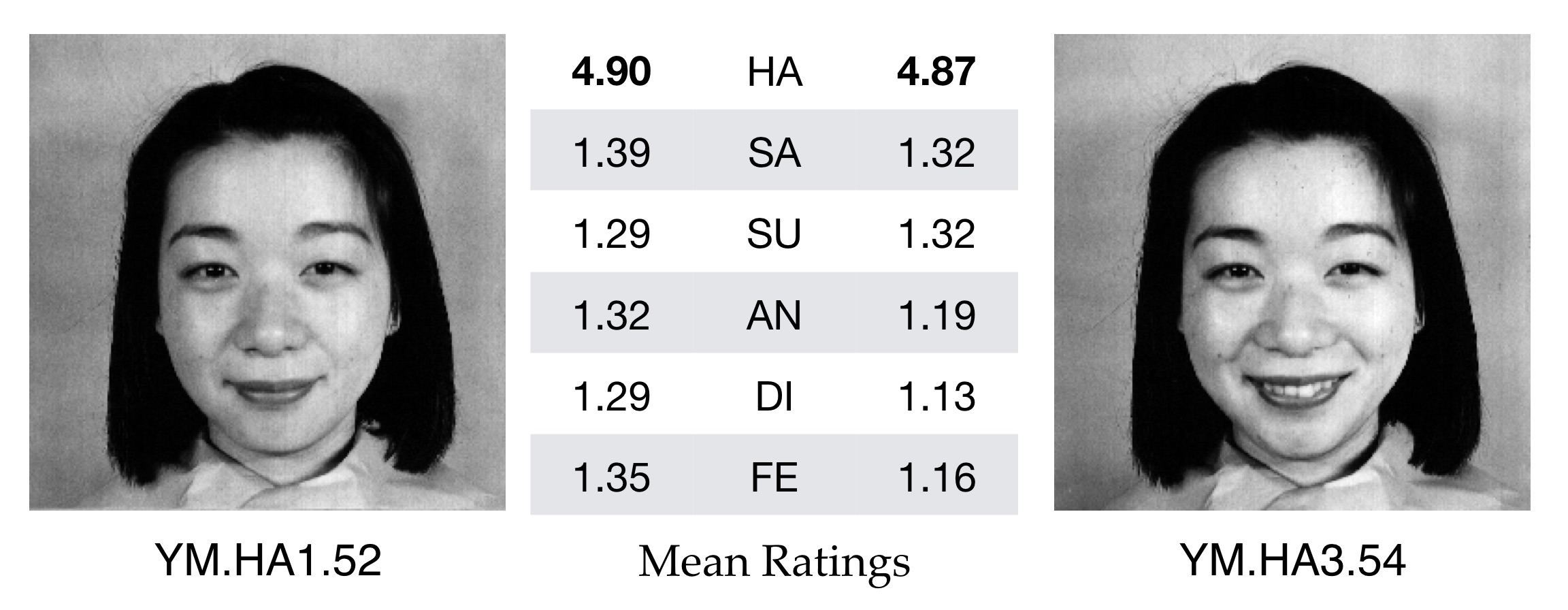}
	\caption{Two happy expressions posed by YM.}
\label{jaffe_ha}
\end{center}
\end{figure}

\begin{figure}[t]
\begin{center}
	\includegraphics[width=1.\linewidth]{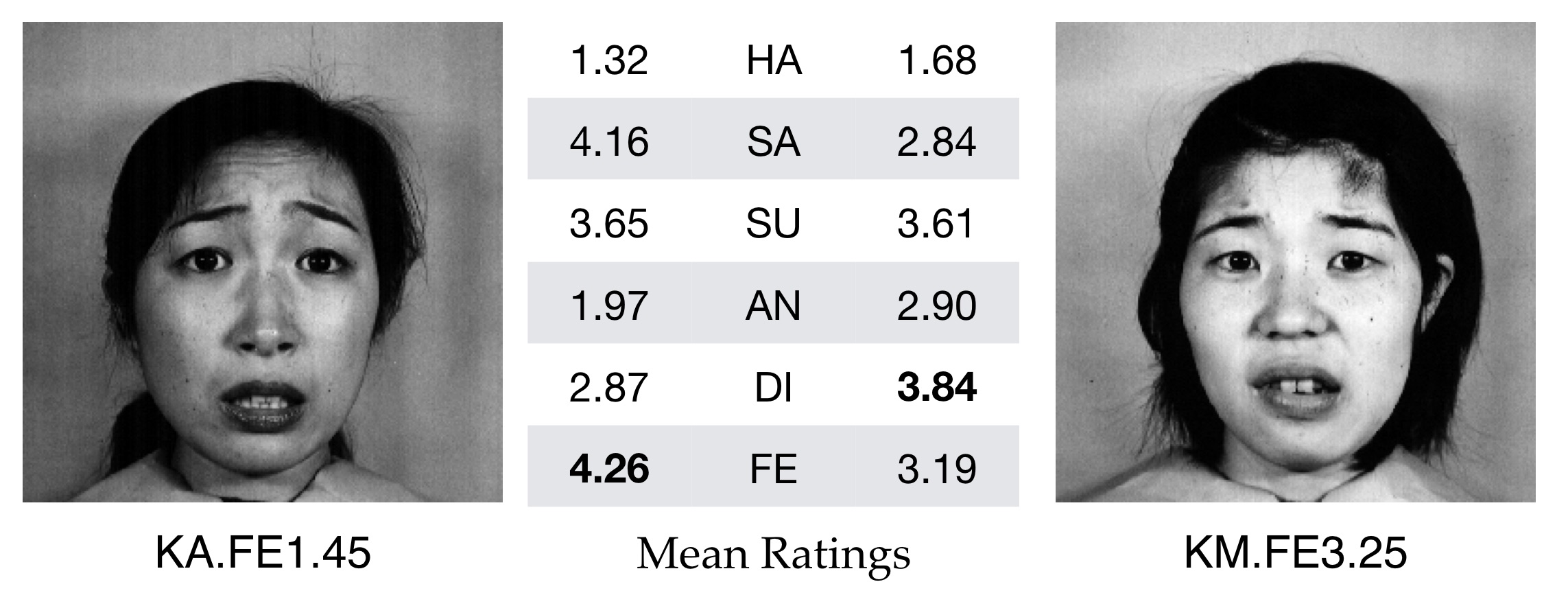}
	\caption{Fear expressions posed by KA and KM.}
\label{jaffe_fe}
\end{center}
\end{figure}

JAFFE volunteers posed the requested expressions while facing the camera through a transparent, semi-reflective plastic pane \cite{lyons1998codingivc} and triggered the shutter remotely while viewing the reflection of their face---the JAFFE images are selfies. Unlike some other facial expression image sets (see, for example, \cite{ekman1976pictures,matsumoto1988japanese}), volunteers were not coached to pose specified facial configurations or to imitate prototypes, nor were they trained in the facial action coding system (FACS) \cite{ekman1978facial}). The volunteers were given a Japanese adjectival noun for each of the six basic facial expressions and asked to photograph three or four poses for each of these while attending to their reflected expressions. JAFFE images did not undergo a selection procedure, as with, for example, Ekman's POFA (Pictures of Facial Affect) set, for which an iterative procedure selected the most prototypical images for each basic facial expression \cite{ekman1976pictures}. JAFFE includes nearly all of the photographs posed by all of the volunteers.\footnote{The publicly available version of JAFFE contains 213 images: ten expressers posing three (or in a few cases four) examples of six expressions and a neutral pose. The original, complete set contains 219 images, but the publicly available set omitted six of the images due to an oversight. There is nothing unusual about the six images missing from the published set.} Because of this, the facial configuration for each expression category varies considerably between poses (figure \ref{jaffe_ha}) and between posers (figure \ref{jaffe_fe}).  

We created JAFFE to experimentally test a mathematical model of facial expression representation derived from a simplified description of the receptive fields of neurons in the visual cortex \cite{lyons1997v1,lyons1996model,lyons1998codingivc,lyons2000linked}. We aimed to compare the model with human perception. To reduce arbitrary assumptions, we compared image similarity derived from the model with similarity measured from human observations of the facial expression images. We did not use categorical facial expression labels directly in the comparison. Instead, we obtained an empirical six-dimensional description for each image by asking observers to view expression images and rate the perceived expression using a five-level Likert scale for each of the six adjectival nouns used to request poses from the volunteers. Images were viewed and rated by 60 Japanese female observers (volunteers recruited at Kyushu University), and the responses averaged. Figure \ref{jaffe_ha} shows sample mean ratings for two JAFFE images. The semantic ratings afford a more detailed and nuanced description of the images than that obtained with the more commonly used forced categorization approach \cite{barrett2019emotional}. Similarity calculated from the multidimensional ratings was correlated with similarity derived from our mathematical model. Model and empirical similarity measures were also analyzed using non-metric multidimensional scaling, revealing low dimensional structure resembling the affective circumplex \cite{russell1980circumplex,schlosberg1952description}.

The approach outlined above does not require the JAFFE expressions to be prototypical. Strictly speaking, neither was it required that the volunteers pose authentic or felt expressions because we compared data on the \emph{observers' perceptions} with the mathematical model, not the poser's \emph{intended expression}. \emph{Category labels are not directly used in our analysis} \cite{lyons1998codingivc}. This does leave open the question of how posed expression images relate to facial expressions encountered `in the wild,' and we were fully aware that our research project did not address this point. 

A large number of scientific studies in the published literature make use of posed expressions \cite{barrett2019emotional}: our study and JAFFE are not unusual in this respect. As it happens, Lisa Feldman Barrett, whom Crawford and Paglen often cite as an Ekman critic, distributes the \emph{IASLab Face Set}, a dataset of posed facial expression images categorized and labelled with emotion nouns.\footnote{The IASLab images and data are available for non-profit academic research at:\\ \url{https://www.affective-science.org/face-set.shtml}.} IASLab does not include multidimensional semantic ratings data: would it not serve better than JAFFE as the `Anatomical Model' for \q{Excavating AI}?

\begin{table}[b]
\centering
\begin{tabular}{|c|c|c|c|c|c|c|c|}
\hline
\multicolumn{2}{|c|}{\multirow{2}{*}{}}                                         & \multicolumn{6}{c|}{RATED} \\ \cline{3-8} 
\multicolumn{2}{|c|}{} & HA  & SA   & SU   & AN   & DI   & FE   \\ \hline
\multirow{6}{*}{\begin{tabular}[c]{@{}c@{}}P\\ O\\ S\\ E\\ D\end{tabular}} & HA & 100\%  &   &   &   &   &   \\ \cline{2-8} 
          & SA         &     & 97\% &      &      & 3\%  &      \\ \cline{2-8} 
          & SU         & 3\% &      & 97\% &      &      &      \\ \cline{2-8} 
          & AN         & 3\% &      &      & 80\% & 17\% &      \\ \cline{2-8} 
          & DI         & 3\% &      &      & 6\%  & 90\% &      \\ \cline{2-8} 
          & FE         &     & 16\% & 31\% &      & 28\% & 25\% \\ \hline
\end{tabular}
\caption{JAFFE posed-rated agreement by facial expression. Agreement is reported here as the percentage of images for which the highest rated expression noun agrees with the requested pose. The average agreement is 81\% over all poses and 93\% if fear poses are not taken into account.}
\label{tab:table1}
\end{table}
\subsection*{Cultural Specificity of JAFFE}
As one might expect, JAFFE exhibits cultural specificity. At the outset of our project, when we were thinking through the design of JAFFE, Jiro Gyoba, an experienced perceptual psychologist, warned that we could not expect Japanese subjects to recognize images of fear poses reliably. Previous studies of facial expression recognition involving Japanese subjects \cite{russell1993canadian} reported such findings. Most experiments have found that negative valence expressions are recognized less reliably in Japan than in Western cultures. This effect is most significant with fear expressions, where poser-observer agreement is measured to be just above chance. 

Nevertheless, we included fear poses in the experiments. Our experimental observations confirmed Jiro's prediction (see Table \ref{tab:table1} and figure \ref{jaffe_fe}): with fear poses, the agreement between the intended expression and the observer ratings is slightly above chance (17\%). Fear poses are rated more strongly as surprise and disgust than fear itself (Table \ref{tab:table1}). These findings, however, did not change the conclusions of our study because we did not force categorization of the facial expressions. In parallel with the measurement on six expression words, we carried out independent measurements from a separate group of observers that did not include fear poses or ratings. Whether or not we included fear poses and ratings, the visual-cortex-inspired model showed a significantly higher correlation with observer ratings than the control, a similarity measure derived from feature-point geometry.

\begin{figure}[t]
\begin{center}
	\includegraphics[width=1.\linewidth]{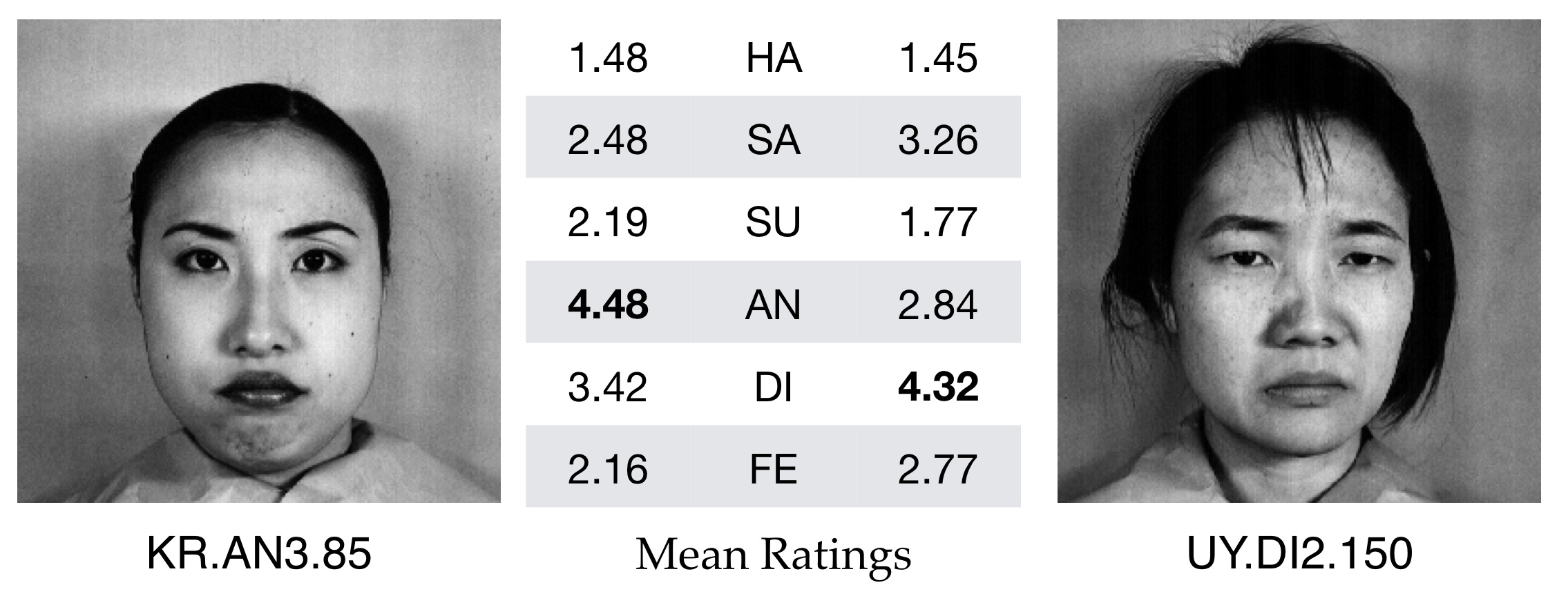}
	\caption{Anger and disgust expressions posed by KR and UY.}
\label{jaffe_an_di}
\end{center}
\end{figure}

We confirmed the cultural specificity of the JAFFE expressions a few years later in a collaborative cross-cultural study of facial expression judgments by Americans and Japanese \cite{dailey2010evidence, dailey2002cultural}. The study found, for example, that the anger pose shown in the left panel of figure \ref{jaffe_an_di} was classified as anger by 82\% of Japanese study participants but only 34\% of American participants. Similarly, the disgust pose shown in the right panel of figure \ref{jaffe_an_di} was classified as disgust by 66\% of Japanese participants but only 18\% of American participants \cite{dailey2010evidence}.

\subsection*{JAFFE as a Training Set}
We first reported the results of our project at the May 1997 edition of \emph{ARVO}\footnote{The \href{https://www.arvo.org/}{Annual Meeting of the Association for Research in Vision and Ophthalmology}.} a vision science conference held in Florida \cite{lyons1997v1}, and at a domestic cognitive science meeting held in Okinawa in June 1997 \cite{lyons1997gabor2}. We learned that the \emph{IEEE Face and Gesture} conference would take place not far from ATR,\footnote{The \href{https://www.atr.jp}{Advanced Telecommunications Research Institute International} (国際電気通信基礎技術研究所) an institute located in Kyoto devoted to basic research, where I worked from 1996-2007.} and prepared an article for submission. The article \cite{lyons1998coding} was shortlisted for a best paper award and invited to be published in a special issue of the Elsevier journal \emph{Image and Vision Computing}.\footnote{The final manuscript that we prepared and submitted for publication in the \emph{Image and Vision Computing} IEEE FG'98 special issue is available at the arXiv open access repository \cite{lyons1998codingivc}. The special issue of \emph{Image and Vision Computing} never materialized.}

After hearing my talk, Zhengyou Zhang, a computer vision researcher who visited ATR in 1997, suggested we extend the visual-cortex-inspired model by training and testing a multi-layer neural network to classify the JAFFE facial expression images. By leveraging existing work, the project progressed quickly, and an article \cite{zhang1998comparison} was submitted to the same edition of the \emph{IEEE Face and Gesture} conference. Zhang compared the classification rates obtained with the neural network directly with the poser-rater agreement levels (see table \ref{tab:table1}) to conclude that the neural network had exceeded human performance. However, this comparison was misleading because the neural network generalization rate measurement had not distinguished individual posers. Moreover, the human observers had rated the facial expression images without any initial instruction on poser intentions, whereas the neural network model was exposed to intended pose labels during training via supervised learning. In other words, the human observers were not `trained' using the pose labels. In accord with my objections, the results of a subsequent study showed that we could equal or slightly exceed the performance of the multilayer neural network using a simple statistical model, while the cross-individual generalization rate was significantly lower \cite{lyons1999automatic}.

In the interests of an \emph{Open Data} policy \cite{murray2008open}, we decided to provide JAFFE images and ratings to others for use in non-commercial scientific research. To my surprise, JAFFE became a frequently used benchmark dataset in academic research on facial expression classification algorithms. By 2007, such studies had become so numerous that I publicly questioned the value of optimizing expression classification on benchmark datasets and cautioned that this was not the same research problem as emotion recognition \cite{lyons2007dimensional}.\footnote{\label{fechner}Invited talk given at Fechner Day 2007, the 23rd Annual Meeting of International Society for Psychophysics held in Tokyo, October 20–23, 2007.}  

Despite the impression given by \q{Excavating AI,} JAFFE is not used exclusively for machine learning research but also in many other fields, including visual psychology, cognitive science, and experimental neuroscience.

\section*{`Archaeology' of JAFFE: The Bones of Contention}
\noindent
Crawford and Paglen adopted JAFFE as their `anatomical model' of machine learning training sets for the essay \q{Excavating AI: The Politics of Images in Machine Learning Training Sets} \cite{crawford2019excavating}. The relevant discussion is in the section titled \q{Anatomy of a Training Set,} that they intend to serve as an exposition of what they describe as their `dataset archaeology methodology.' 

Nearly every statement about JAFFE in this important theoretical section of \q{Excavating AI} \emph{is unequivocally false}. In the following, excerpts from Crawford and Paglen's account of JAFFE are quoted verbatim, in italics. My commentary, in plain font, follows each excerpt. 
The excerpts are quoted from the web version \cite{crawford2019excavating}. The text of the AI \& Society article is nearly identical.

\subsection*{\emph{Emotional States}}
\begin{quote}
\textit{
The dataset contains photographs of 10 Japanese female models making seven facial expressions that are meant to correlate with seven basic emotional states.
}
\end{quote}

\noindent We asked volunteers to freely pose six basic facial expressions as well as an expressionless face. We did not instruct or guide the volunteers to imitate specific prototypical facial action configurations. The volunteers were not `models.'

\addcontentsline{toc}{subsection}{Intended purpose}
\subsection*{\emph{Intended purpose}}
\begin{quote}
\textit{
The intended purpose of the dataset is to help machine-learning systems recognize and label these emotions for newly captured, unlabelled images.
}
\end{quote}
\noindent JAFFE was designed and photographed to compare a mathematical model of facial expression representation with data from human observers \cite{lyons1998coding,lyons1997gabor2,lyons1997v1,lyons1998codingivc}. Our project did not train a machine learning system, nor did it 'label emotions for newly captured images.' A central concern of \q{Excavating AI} is the critique of image taxonomy; ironically, Crawford and Paglen's mistaken description of JAFFE's origins as a machine-learning training set is itself a taxonomic error.  

\subsection*{\emph{Depicting Emotions}}
\begin{quote}
\textit{
The implicit, top-level taxonomy here is something like \q{facial expressions depicting the emotions of Japanese women.}
}
\end{quote}
We have never described JAFFE as anything but a set of images of posed facial expressions. The accompanying article \cite{lyons1998coding, lyons1998codingivc} describes in detail the procedure for photographing the JAFFE images. We made no explicit or implicit claim that these depict felt emotions.

\addcontentsline{toc}{subsection}{Organizing Buckets}
\subsection*{\emph{Organizing Buckets}}
\begin{quote}
\textit{
If we go down a level from taxonomy, we arrive at the level of the class. In the case of JAFFE, those classes are happiness, sadness, surprise, disgust, fear, anger, and neutral. These categories become the organizing buckets into which all of the individual images are stored. They are the distinct concepts used to order the underlying images.
}
\end{quote}
The title of each JAFFE image file is composed of a two-letter volunteer identifier, two-letter pose identifier, pose number, and image number. The distribution includes a README\_FIRST file containing multidimensional ratings for each image based on sixty Japanese observers' judgments. README\_FIRST contains the following explicit caveat:
\begin{quote}
$\ldots$ it is important to realize that the expressions are never pure expressions of one emotion but always admixtures of different emotions. The image expression labels represent the predominant expression in that image\textemdash the expression that we asked the subject to pose.
\end{quote}
\noindent Intended pose may not agree with the observer ratings, as in the right panel of figure \ref{jaffe_fe}. Crawford and Paglen's concept of categorical `organizing buckets' ignores the nuanced description available from the subjective rating data and is a misleading characterization of JAFFE.

\subsection*{\emph{Emotions as Visual Concepts}}
\begin{quote}
\textit{
There are several implicit assertions in the JAFFE set. First there’s the taxonomy itself: that \q{emotions} is a valid set of visual concepts. 
}
\end{quote}
JAFFE makes no such assertion, explicitly or implicitly. Our study modelled ratings by observers, not the emotions or mental states of those posing the expressions. Uninformed users of JAFFE might assume this about the image labels if they have not read and understood README\_FIRST and the article that describes JAFFE \cite{lyons1998coding}.

\addcontentsline{toc}{subsection}{Emotions applied to Photographs}
\subsection*{\emph{Emotions applied to Photographs}}
\begin{quote}
\textit{
Then there’s a string of additional assumptions: that the concepts within \q{emotions} can be applied to photographs of people’s faces specifically Japanese women) \ldots
}
\end{quote}
\indent We made no such claims about the emotions of the volunteers who posed the expressions. Our study modelled perceived expression based on observer ratings, not felt emotion. Crawford and Paglen have misunderstood this point. Our measurement protocol (semantic rating on expression nouns) is not unusual but widely used---the misunderstanding is symptomatic of a lack of basic familiarity with the published literature on the psychology of facial expression.

\subsection*{\emph{Six Emotions}}
\begin{quote}
\textit{
\ldots that there are six emotions plus a neutral state \ldots 
}
\end{quote}
\noindent Our work is concerned with the visual perception of facial expressions, not emotions of the posers. We have never stated that there are only six emotions or facial expressions, a nonsensical claim. 
\addcontentsline{toc}{subsection}{True Emotional State}
\subsection*{\emph{True Emotional State}}
\begin{quote}
\textit{
... there is a fixed relationship between a person’s facial expression and her true emotional state ... 
}
\end{quote}
\noindent This premise has nothing to do with JAFFE, which is instead characterized by subjective ratings data and accompanied by our explicit caveat about the meaning of the pose labels. 

\subsection*{\emph{Consistent and Uniform}}
\begin{quote}
\textit{
\ldots and that this relationship between the face and the emotion is consistent, measurable, and uniform across the women in the photographs.
}
\end{quote}
\noindent This premise cannot derive from actual observation of JAFFE images and data. Look again at the image samples and ratings in figures \ref{jaffe}, \ref{jaffe_ha}, \ref{jaffe_fe}, and \ref{jaffe_an_di}. Each volunteer posed several variations for each facial expression. No two images in the JAFFE set have identical facial expressions or rating values. 

\subsection*{\emph{Neutral Facial Expression}}
\begin{quote}
\textit{
At the level of the class, we find assumptions such as “there is such a thing as a \q{neutral} facial expression
}
\end{quote}
\noindent While the concept of a neutral facial expression is not meaningless, the ratings data indicate that none of the images depicting expressionless (無表情) poses were perceived by the observers as devoid of expression.
\addcontentsline{toc}{subsection}{Significant Six Emotional States}
\subsection*{\emph{Significant Six Emotional States}}
\begin{quote}
\textit{
and “the significant six emotional states are happy, sad, angry, disgusted, afraid, surprised.”
}
\end{quote}
\noindent Crawford and Paglen have failed to understand a fundamental aspect of our research project: it is concerned with the \emph{visual appearance of facial expressions} in images, not 'emotional states.' Regarding the significance of the facial expression categories chosen for the study, these are, of course, the \emph{Basic Facial Expressions} (BFEs)  used in many psychological studies of facial expression, not only by the Ekman camp. The use of these expression categories in research dates back at least to Charles Darwin's work \cite{darwin1872express}. Our project did not assume that these are the \emph{only} expressions, nor did we make any such claim.  We were well aware of criticism of BFE theory. Indeed one of the aims of our work was to compare categorical and dimensional descriptions of facial expressions \cite{lyons2020excavating, lyons1998codingivc}. Furthermore, our project did not rely on the validity of the BFE paradigm. 

\subsection*{\emph{Interior State}}
\begin{quote}
\textit{
At the level of labelled image, there are other implicit assumptions such as \q{this particular photograph depicts a woman with an ‘angry’ facial expression,} rather than, for example, the fact that this is an image of a woman mimicking an angry expression. These, of course, are all \q{performed} expressions—not relating to any interior state, but acted out in a laboratory setting. 
}
\end{quote}
One wonders if Crawford and Paglen have looked at \emph{any} of the documentation associated with JAFFE. We clearly describe the images as depicting \emph{posed facial expressions}. Our article clearly describes the laboratory conditions under which we photographed JAFFE \cite{lyons1998coding}. There is no claim or implication that JAFFE images depict natural expressions, and there is no discussion of `interior state.'

\subsection*{\emph{Implicit Claims}}
\begin{quote}
\textit{
Every one of the implicit claims made at each level is, at best, open to question, and some are deeply contested.
}
\end{quote}
\noindent An implication is a compound statement having the form \emph{X implies Y.} But Crawford and Paglen's statements concerning JAFFE take the form `\emph{Y is implied},' without specifying a premise, \emph{X}, and without providing the slightest explanation about how these unknown factors imply the alleged `claims.'
All of the allegations about JAFFE's `implicit claims' are furnished as self-evident conclusions made without evidence, elaboration, or reference to statements in the relevant publications.

\subsection*{\emph{Relative Modesty}}
\begin{quote}
\textit{
The JAFFE training set is relatively modest as far as contemporary training sets go. It was created before the advent of social media, before developers were able to scrape images from the internet at scale, 
}
\end{quote}
\noindent Crawford and Paglen add to the fiction that we intended JAFFE as a training set. The fabulation continues: JAFFE is a scale-challenged `modest' training set, stunted by the impossibility of scraping for images. Images scraped online would not have been suitable for the project. JAFFE was designed and explicitly photographed for our experiment.  The relatively small size of the JAFFE set had nothing to do with a lack of big data: we chose the numbers of volunteers and poses to permit a statistically significant comparison of two measures of facial expression similarity \cite{lyons1998codingivc}. Had we wished to design an image dataset for facial expression machine learning, we would have recruited more volunteers. 

\section*{The Straw Men}
Each of the alleged `implicit claims' about JAFFE is a misrepresentation of our work. As a member of the team that created JAFFE, I can state unequivocally that these allegations do not correspond with our views or intentions. 

How did Crawford and Paglen arrive at such a mistaken account of our work?  It appears that the veracity of their allegations about JAFFE was not a priority. Instead, these serve a mainly rhetorical function. Examining the list of  statements reveals a pattern:
\begin{itemize}
	\item Each statement is made without evidence (e.g.\ `JAFFE is intended as a machine learning training set')
	\item Each of the `implied claims' is easy to refute without specialized knowledge (e.g.\ `posed images depict interior states')
	\item Some of the `implied claims' are ridiculous (e.g.\ `there are only six emotions')
\end{itemize}

A \emph{straw man} argument is a common fallacy of informal reasoning that distorts or misrepresents a position in order to make it easier to refute \cite{walton2008informal}. This fallacy provides the key to understanding Crawford and Paglen's account of our work. As we have seen, \q{Excavating AI} does not critique factual attributes of JAFFE, but nonfactual `implied claims'---misrepresentations posited without a shred of evidence. Each false allegation is simplistic and intuitively refutable. Some are ridiculous: what kind of deranged charlatan would propose that there are no more than `six facial expressions and a neutral face' and that it is possible to 'mind read' from a single photograph of a posed expression? These ridiculous caricatures, fabulated by Crawford and Paglen in \q{Excavating AI,} appear entirely believable by readers who have limited knowledge of the research literature on the psychology of facial expression. 

The straw man offers several incentives: it is easier to communicate simplistic, fallacious views to a non-specialist reader than faithful, nuanced ones. Imagined `claims' can be constructed for ease of refutation. Distorted caricatures are more memorable than plain reality. An influential study has found that falsehood spreads more quickly on social networks than the truth \cite{vosoughi2018spread}.  Another recent study found that populist politicians employ informal fallacies to attract attention and persuade \cite{blassnig2019populism}. Does \q{Excavating AI} frame JAFFE using straw man arguments for rhetorical purposes? This is the most plausible explanation I have found for the consistently erroneous account of JAFFE given in the essay. 

\subsection*{Cognitive Biases}
\q{Excavating AI} has enjoyed a largely positive reception by humanities scholars who cite it as having identified problematic epistemology in machine learning training sets. We do not deny the importance of this issue---there is no doubt that there are machine learning datasets embodying questionable epistemology. Nonetheless, it is surprising that the incorrect description of JAFFE has slipped past the discernment of so many: how can the fallacies have escaped the critical faculties of these readers? 

First, consider the context in which \q{Excavating AI} has appeared: the decade preceding its publication has witnessed accelerated interest and investment in artificial intelligence and machine learning and the intensification of ideology that favours technology as the primary driver of social and economic innovation. Resistance has also grown: increasingly, the critical study of technology is a valued scholarly activity that is impacting public discourse and policy. Readers of \q{Excavating AI} who are skeptical of AI hype may be susceptible to \emph{confirmation bias} \cite{colman2015dictionary} that leads them to overlook invalid parts of Crawford and Paglen's argument unwittingly.

The \emph{halo effect} \cite{colman2015dictionary} could be another factor: Crawford and Paglen are frequent, high-profile speakers at public events, sometimes portrayed in the media as heroic social activists: we can expect a level of positive bias. 

\emph{Narrative bias} \cite{borgida1977differential} may also be a factor: \q{Excavating AI} offers readers a compelling narrative that can overrule judgments on the veracity of its elements. Readers may assume that the `implied claims' linked to JAFFE are factual, even without being offered evidence, because this interpretation fits the essay's overall theme: false image taxonomy. Ironically, Ren\'{e} Magritte's \emph{La Trahison des Images}, a point of reference in Crawford and Paglen's thesis, is relevant here: a false ontology accompanies the JAFFE images---the straw man arguments offer the reader/viewer a deceptive \emph{misrepresentation}. 

Crawford and Paglen provoked a media spectacle involving: the viral selfie app \emph{ImageNet Roulette}; the highly visible art exhibitions \emph{Training Humans} and \emph{Making Faces} sponsored by global fashion brand \emph{Prada}; mass media exposure that included coverage by the Netflix tech-lite variety show \emph{Connections}; and torrents of activity on the social networks \cite{lyons2020excavating,tedone2019from}. Public attention accrued to such a degree that the \emph{bandwagon effect} \cite{colman2015dictionary} may significantly bias consumers' judgments, even if widespread popularity is no evidence for validity.

To recognize the misinformation, a reader would need to examine and understand the documentation accompanying JAFFE. It is unlikely that most readers will go to such lengths. Taken together with the biases I have outlined, this helps us understand how the fallacies concerning JAFFE pass inconspicuously.

\section*{Mind Reading Machines?}
One of the most striking of the `implied claims' in \q{Excavating AI} concerns the purported ability to read  `true emotional state' or `interior state' from images of posed facial expressions. As I have shown, our work made no claim to read internal states. Where did Crawford and Paglen get such ideas? Perhaps their imaginations were stimulated by media reports of emotion reading software and the websites of affective computing startups? 

\noindent Here are some press headlines and excerpts from company web pages:
\begin{itemize}
	\item \emph{Apple Buys Emotient, a Company That Uses AI to Read Emotions} \cite{byford2016apple}
	\item \emph{\ldots Startup That Can Read Your Face To Tell If You're Angry} \cite{solomon2016apple}
    \item \emph{AI Can Read Your Emotions. Should it?} \cite{lewis2019ai}
	\item \emph{Human Perception AI. Our software detects all things human: nuanced emotions, complex cognitive states\ldots}\footnote{\emph{Affectiva} homepage: \url{https://www.affectiva.com/} (accessed July 12, 2021).}
\end{itemize}

\noindent My first encounter with this genre of publicity dates to about 2006, when the New York Times reported \cite{schuessler2006social} on a project by Rana el Kaliouby and Rosalind Picard of the MIT Media Lab, who would found the startup \emph{Affectiva} a few years later \cite{el2020girl}. The article described an \q{Emotional-Social Intelligence Prosthesis} that promised to augment an autistic user's ability to `mind read' by providing facial expression recognition support. Some might na\"{i}vely misinterpret the mention of \emph{mind reading} as an `implied claim' hinting at telepathy or phrenology. It is nothing of the kind: the project and el Kaliouby's earlier Ph.D. thesis \cite{el2005mind} drew on the \emph{Theory of Mind} theory of cognitive science---the hypothesis that we use folk psychology to make inferences about the beliefs and intentions of other beings \cite{baron1985does, churchland1986neurophilosophy, dennett1987intentional, premack1978does}.  The New York Times report mentioned Simon Baron-Cohen, who proposed a model of autism based on the \emph{Theory of Mind} theory \cite{baron1997mindblindness, baron1985does} and was one of el Kaliouby's thesis advisors. 

I was impressed by el Kaliouby's ambitious attempt to model facial expression perception more meaningfully than the image classification studies still commonplace.  At the same time, I was uncomfortable with the anthropomorphic description of computer software. I expressed my reservations at conferences in Tokyo in 2007 \cite{lyons2007dimensional}\footnote{The Fechner Day, 2007 conference mentioned in footnote \ref{fechner}.} and in Amsterdam in 2008 \cite{lyons2008mimetic}.\footnote{IEEE Face and Gesture Conference, Workshop on Facial and Bodily Expressions for Control and Adaptation of Games, Amsterdam, NL, September 16, 2008.} My skepticism was twofold. First, as mentioned above, the term `mind reading' has alternative quasi-magical connotations likely to mislead the popular imagination. The researchers could have avoided potential misunderstandings by describing their work as, for example, a computational model of the theory of mind, which would be accurate though less sensational. Second, there were practical difficulties involved in interpreting facial displays. My presentations in Tokyo and Amsterdam gave a simple demonstration\footnote{See the Amsterdam presentation slides: \url{https://zenodo.org/record/4026982}.} showing that the crucial role played by context would be difficult to take into account generally. Barring a solution of the hard AI problem, one did not expect `mind reading' systems to function except in rather constrained situations. Indeed, as late as 2011, after several years of further development, \emph{Affectiva}'s technology was still having difficulty distinguishing some types of grimace from smiles  \cite{el2020girl}.  

\emph{Affectiva} evolved and is valued by investors and clients in the automotive, marketing, and advertising industries, but Shoshana Zuboff argues that, despite initial good intentions, \emph{Affectiva} was drawn into \q{surveillance capitalism's powerful force field.}\footnote{Chapter 9, section III, \q{Machine Emotion,} of \emph{The Age of Surveillance Capitalism} \cite{zuboff2019age}.} I wonder if the outcome would have been more benign had el Kaliouby continued to pursue the computational modelling of the \emph{Theory of Mind} theory scientifically rather than commercially?

\section*{Challenging the Common View}
In December 2000, I took part in a workshop on affective computing held as part of the Neural Information Processing Systems conference (aka NeurIPS).\footnote{\url{https://www.cs.cmu.edu/Groups/NIPS/NIPS2000/Workshops/}} The call for participation encouraged speakers
\begin{quote}
\textit{
$\ldots$ to talk about challenges and controversial topics both in their prepared talks and in the ensuing discussions.\footnote{\url{https://www.kdnuggets.com/news/2000/n20/34i.html}} 
}
\end{quote}
My contribution offered two challenges and a proposal \cite{lyons2000nips}.\footnote{NeurIPS'2000 Workshop presentation slides: \url{https://zenodo.org/record/5016197}.}

My first challenge to the workshop reported our experimental finding that head tilt influences facial expression perception. Our cross-cultural study of this phenomenon, in collaboration with Ruth Campbell at the University College London, was inspired by the variable expressions of masks used in the Japanese Noh theatre \cite{lyons2000noh}. Tilting the mask toward or away from a viewer changed the perceived expression. The effect differed for Japanese and British viewers. Our experiments demonstrated that a related effect obtains with human faces. This revealed a challenge: most research on facial expression recognition had assumed frontal, untilted views of the face. If slight changes in head pose could change the apparent expression, how had this influenced psychological studies of expression perception? Moreover, our results showed clearly that automatic facial expression recognition algorithms would need to account for head pose variations.

My second challenge concerned the cultural variability of facial expressions. Our study of Noh masks had revealed a cultural difference between Japanese and British experimental subjects. Furthermore, we had already observed cultural specificity with the JAFFE images and ratings data \cite{lyons1998codingivc}. I recommended that affective computing researchers should carefully re-examine the facial expression universality hypothesis.

Next, I proposed an alternative paradigm for facial expression computing, suggesting that it would be worthwhile to pursue the development of systems permitting the use of \emph{voluntary} facial gestures for \emph{intentional} human-computer interaction. I had already been working on such \emph{facial gesture interfaces} \cite{lyons2004facial}. To illustrate the concept, I showed workshop attendees a video of my \emph{mouthesizer} prototype: a wearable facial gesture interface allowing musical performers to control audio effects in real-time with their mouths \cite{lyons2001facing}. 

My presentation concluded with the announcement of a workshop we were organizing at the upcoming ACM CHI'2001 conference\footnote{ACM Conference on Human Factors in Computing Systems, March 31 - April 5, 2001.} in Seattle, entitled \q{New Interfaces for Musical Expression,} and encouraged interested listeners to get involved. The NIME workshop would explore prospects for interdisciplinary research involving human-computer interaction and musical expression \cite{poupyrev2001new}.

I wonder what impact, if any, the unconventional presentation had on the workshop attendees. Several years later, three of the four workshop organizers founded the startups \emph{Affectiva} and \emph{Emotient} to commercialize technology to recognize `emotion' displayed involuntarily by the face. In the audience was the director of the US Department of Defence sponsored Facial Recognition Technology (FERET) program.\footnote{\url{https://www.nist.gov/programs-projects/face-recognition-technology-feret}} He had organized competitive tests of automatic facial recognition algorithms and told me he was now looking into possibilities in the affective computing field. I recall asking the other speakers during the coffee break if they were not concerned about the potential exploitation of their research for surveillance purposes. 

My presentation had at least one tangible outcome: after the workshop, discussion continued with the fourth workshop organizer, cognitive scientist Gary Cottrell: it was the beginning of our collaborative study of culturally dependent aspects of facial expression perception \cite{dailey2010evidence,dailey2002cultural}.

The NeurIPS'2000 Affective Computing workshop is a personally significant landmark that signalled a shift in my research activities: I had already begun to prioritize the voluntary, intentional, and expressive aspects of action and enaction. It also marked the gestation of a new research community, NIME,\footnote{NIME Community Portal: \url{https://www.nime.org/}.} that would grow and confer every year after the first CHI'2001 workshop. NIME  offers a leading venue for scientific research and artistic practice at the intersection of musical expression and technology \cite{jensenius2017nime}. In December 2000, NIME extended the possibility of a new locus of positive resistance against the infiltration of the surveillance paradigm into the human-machine relationship and an exploration of the potential of technology in the realm of artistic expression.

\section*{Algorithms of Inequity}
If Crawford and Paglen's essay is an excavation, then JAFFE is its Piltdown Man, a hoax created by would-be archaeologists \cite{lewin1997bones}. As we have seen, the JAFFE dataset was abused as fodder for rhetorical straw man arguments about machine learning training data. But what can we conclude about the essay in general?  Is it not advisable to scrutinize other aspects of \q{Excavating AI}? 

Consider the overall focus on training set taxonomy. Imagine, for purposes of argument, that collaboration between socially aware engineers and humanities scholars has managed to resolve the data taxonomy issues raised in \q{Excavating AI.} Would this lead to the development of ethical facial recognition technology?  Should the possibility of reducing bias ease concerns about the deployment of surveillance systems? Will dataset debiasing efforts by the surveillance capitalist behemoths amount to anything more than \emph{digital ethics bluewashing} \cite{floridi2019translating}?

Whether for policing or profit, surveillance technology affords mediated social relations that are, by definition, unequal.\footnote{From the French, the verb \emph{surveiller} adds the prefix \emph{sur-} (over) to \emph{veiller} (to watch) meaning `to over watch' implying unequal vigilance.} Automated surveillance is a species of machine-mediated interaction for which the framing of \emph{interaction} as an \emph{exchange of information} lacking \emph{detailed reciprocity} leads directly to algorithms for the generation of inequity. Taxonomy fixes will never repair the root issue: surveillance presupposes a bias that privileges the watcher over the watched. By contrast, ethical human-machine systems afford transparency, volition, and reciprocity. Such systems do not engender or enhance power asymmetries and social inequity. I embraced these views in the 1990s, when a scientific interest in face perception briefly brought me into contact with the field of surveillance technology.

\section*{Lessons Learned?}
I want to share some of the insights gained from debunking Crawford and Paglen's fallacious account of the JAFFE dataset. These lessons may seem self-evident but will be stated explicitly in the general spirit of this commentary.

Foremost, before criticizing something, it is necessary to understand it correctly. If there is a README file, read it. If there is associated documentation, study it. Failing to do this, one generates misinformation and misleads readers. 

If anything is unclear in the documentation, and even if there is not, get in touch with the authors. Find out what they think about your analysis. 

If a target of your critique raises objections, listen to what they have to say and understand their viewpoint. To do so is both good research practice and a matter of basic respect for other human beings.

If some part of your critique is mistaken, admit it. Sooner is better than later: avoid escalating a commitment to beliefs that have already been proven to be incorrect.

Before using facial images for an art exhibition, or publicity, clearly understand the terms of use and confirm the informed consent of the persons depicted \cite{lyons2020excavating}. The pamphlet entitled \q{Making AI Art Responsibly: A Field Guide} by the Partnership on AI \cite{saltz2020making}\footnote{\url{https://www.partnershiponai.org/wp-content/uploads/2020/09/Partnership-on-AI-AI-Art-Field-Guide.pdf}} offers good advice about reusing image data and other resources. Do not misjudge the informal presentation style: the guide is well researched and based on a solid understanding of its subject.

Finally, to critique affective computing technology, choose a worthy target rather than a convenient straw man. The leading companies in the field began as research projects that spawned startups: paper trails of academic publications and patent applications are readily available. Less well documented approaches should be amenable to fieldwork and some reverse engineering. The recently published investigation of the \emph{Vibraimage} system is
an excellent example of thorough and productive critical research \cite{wright2021suspect}. 

\vspace{5 mm}
{\noindent\small \textbf{Michael Lyons is Professor of Image Arts and Science at Ritsumeikan University}}

\vspace{3 mm}


\nocite{muller2012trans}
\bibliographystyle{acm}
\bibliography{eair.bib}

\end{CJK}
\end{document}